# A hydrogen-informed Rice-Beltz model for crack-tip dislocation emission under mixed-mode loading

Kai Zhao[1,2*]

*1. School of Mechanical Engineering, Jiangnan University, Wuxi 214122, China*

*2. Jiangsu Key Laboratory of Advanced Food Manufacturing Equipment and Technology, Wuxi 214122, China*

## Abstract

The fundamental competition between crack-tip cleavage and dislocation emission dictates the ductile-to-brittle transition in crystalline solids. In hydrogen-charged environments, this delicate balance is disrupted, yet existing models often struggle to self-consistently capture the atomic-scale thermomechanical modulation of the emission barrier. Here, we propose the hydrogen-informed Rice-Beltz framework to quantify the critical stress intensity factors (SIFs) required for initial dislocation nucleation under mixed-mode (I+II) loading. By integrating the elastic interaction of the dilatational field of interstitial hydrogen and the applied stress fields against the incipient dislocation core, the model recovers the non-monotonic relation between the most probable SIF and the hydrogen concentration. Unlike the classical pure mode-I (or mode-II) scenario, the minimization of the local strain energy density is employed to generate closed-form nominal driving forces, which are subsequently embedded into a transition-state-theory framework. This atomistically-informed approach yields the most probable SIF for the first dislocation emission event as a function of the loading rate, slip angle, and hydrogen concentration, providing a rigorous, parameter-free boundary condition for the onset of hydrogen-modulated crack-tip plasticity.

***Keywords**:* Hydrogen embrittlement; Rice-Beltz model; Dislocation; Unstable stacking fault energy.

## 1. Introduction

The ductile-to-brittle transition (***DBT***) [1, 2] represents a fundamental shift in the fracture behavior of crystalline materials, characterized by a transition from brittle cleavage at low temperatures to ductile yielding at elevated temperatures. Although observed for more than a century [2], the underlying micromechanical mechanisms governing the ***DBT*** remain a subject of active controversy, specifically, whether the macroscopic transition is predominantly dictated by dislocation mobility [3-5] or the initial dislocation nucleation [6-8] at the crack tip. Within the conventional understanding of intrinsic fracture, the ultimate failure mode is determined by a kinetic competition, *i.e.*, either the crack propagates via atomic bond rupture, or it is blunted by the emission of dislocations [9]. By incorporating the Peierls-Nabarro concept into the description of dislocation nucleation, the Rice-Beltz framework established a rigorous energetic criterion, demonstrating that this emission process is fundamentally controlled by the unstable stacking fault (***USF***) energy ($\gamma_{usf}$). This Peierls-Rice-Beltz (***PRB***) approach has since provided the theoretical benchmark for understanding intrinsic fracture in pristine lattices, elegantly bridging discrete atomic lattice resistance with continuum stress intensity factors.

[*] Corresponding author: kai.zhao@jiangnan.edu.cn (K. Z.)

However, the introduction of hydrogen into the metallic lattice drastically disrupts this inherent stability, rendering the classical ***DBT*** criteria insufficient [10]. Hydrogen embrittlement (***HE***) is not merely a passive alteration of material properties, but a dynamic, multi-scale modulation of the crack-tip field. Upon entering the lattice, hydrogen atoms migrate to regions of high hydrostatic stress, such as crack tips and dislocation cores, where they form localized solute atmospheres [11]. This localized enrichment thermomechanically modulates the intrinsic energy landscape and alters the lattice resistance to shear, effectively tilting the kinetic competition in favor of brittle cleavage. While existing ***HE*** models, *e.g.*, the hydrogen-enhanced decohesion (***HEDE***) [12, 13] and hydrogen-enhanced localized plasticity (***HELP***) [14, 15], often account for this by employing empirical shifts in fracture energy or phenomenological reductions in yield stress, they lack the constitutive depth to describe the multiscale coupling between hydrogen-solute potentials and the dislocation emission barrier, for which I would discuss on three different levels hierarchically.

On the dislocation core width level, the Peierls-Nabarro (***PN***) model [16-19] and its modifications [20-23] have been widely applied to describe singularity fields around dislocation. Although numerous simulations [24-27] have been dedicated to describe dislocation core structures under hydrogen environment, continuum models have not considered the effect of hydrogen on dislocation core energies yet [28]. Grand canonical Monte Carlo (GCMC) simulations conducted by Yu *et al*. [25] reveal that hydrogen increases the core radii and decreases the core energies, leading to the reduction of dislocation line energy. Based on the density functional theory, Li *et al*. [24] revealed a hydrogen-induced transition of screw dislocation core structure in bcc W. Specifically, at low H concentrations, dislocation maintains the intrinsic easy-core structure, and H atoms are attached to the "periphery" of dislocation to enhance dislocation motion; in contrast, at high H concentrations, dislocation transforms into a hard-core, metal hydride-like structure, while H atoms are implanted into the "body" of dislocation to significantly reduce the dislocation mobility. Leyson *et al*. [29] developed a multiscale approach to study the interaction of hydrogen with the core and the strain field of edge dislocations, and revealed a sharp transition between hydride forming and non-hydride forming regimes, with respect to a critical hydrogen chemical potential $\mu_H^c$ related to the nano-hydride nucleus of the system. Recently, Zhang *et al*. [30] proposed a stochastic Peierls-Nabarro (***sPN***) model to account for the standard deviation of the perturbation in the interplanar potential of high-entropy alloys (HEA), leading to stochastic variations in the dislocation width, Peierls stress and Peierls energy. Considering the similarity between the compositional randomness in an HEA and the stochastic distribution of hydrogen atoms within the core width, one might ask whether the ***sPN*** model can be applied for the hydrogen/dislocation interaction, or not?

On the level of short-range interactions, *i.e*., how hydrogen atoms (and derivative hydrogen-vacancy complexes) affect the nucleation [31, 32], and motion of a dislocation, and interactions between two neighboring dislocations [33], *i.e*., the hydrogen-induced "elastic shielding" results in a decrease in the elastic force between two dislocations. As proved by nanoindentation experiments [32, 34], the "defactants" concept [35-39] can be applied to understand the hydrogen effect on dislocation nucleation by reducing the formation energy of dislocations and stacking faults. Leyson *et al*. [40] developed a multiscale model to describe the onset of homogeneous nucleation of a dislocation loop with the presence of hydrogen, with faithfully predicting the experimentally observed drop of the pop-in force. While most studies favor to believe that hydrogen atoms (or hydrogenated vacancies) either enhance [41-44] or impede [45-47] dislocation motion solely, recent *in-situ* scanning electron microscopy experiments [48] reveal that hydrogen can both move or pin dislocation in bcc metals. The kinetic Monte Carlo simulations by Katzarov *et al*. [49] also demonstrate that the steady-state dislocation velocity does not change with the bulk hydrogen concentration monotonically, thus support the dual role effects of hydrogen on the dislocation mobility [50].

On the level of long-range interactions, *i.e*., how hydrogen atoms affect the collective behavior of dislocation ensembles (as well as other types of defects, e.g., grain boundaries (GBs), twin boundaries [51], precipitates [52, 53], vacancy clusters [54]), the underlying mechanisms are still not clear due to the unpredictability of the complex many-body system. While most studies are concerned about the hydrogen trapping capacity of dislocations [55], GBs [56, 57], and precipitates [58-63], little attention is paid to resolve how these preexisting defects influence the plastic evolution under hydrogen environment [64-67]. Combining the *in-situ* electron channeling contrast imaging experiments, molecular dynamics (MD) and GCMC simulations, Koyama *et al*. [68] revealed that the driving force for dislocation movement is provided by the high stress concentration arising from hydrogen segregation at GBs. Based on the mean-field rate theory [69] and grouping numerical method [70], Li *et al*. [54] developed a cluster dynamics model to study the long-time evolution of hydrogen/vacancy system. However, in their model, the dislocation density was set as constant.

To resolve these multiscale debates and bridge the phenomenological gap, one must anchor the analysis at the absolute origin of plasticity, *i.e*., the first dislocation emission event. Starting from the ***PRB*** theory [8], here I attempt to construct a thermomechanically-consistent framework to rigorously describe the dislocation emission from the crack-tip under mixed-mode loading. By decoupling the initial nucleation event from subsequent complex dislocation-forest interactions, this study isolates the precise role of local hydrogen enrichment in triggering the onset of crack-tip plasticity.

## 2. Theoretical framework

In this section, the dislocation emission from the crack-tip under mixed-mode (I+II) loading is derived from the classical ***Rice-Beltz*** theory in **Section 2.1**, by introducing the minimization of the strain energy density to renormalize the localized $K_I^*$ and $K_{II}^*$ fields. Then, the MD settings for evaluating the ***USF*** energy as a function of hydrogen concentration are present in **Section 2.2**.

### *2.1 Hydrogen-informed PRB framework*

Based on the ***Rice-Thomson*** model [6] considering the dislocation shielding effect, the total force applied on the dislocation near the crack-tip under hydrogen environment is evaluated for the configuration shown in **Fig. 1**,

$$f_{tot} = f_K + f_{d^2} + f_{dd'} + f_{gbe} + f_H \, , \tag{1}$$

where, the first term $f_K$ arises from the stress field of the blunt crack-tip, the second term $f_{d^2}$ accounts for the contribution of image dislocations, the third term $f_{dd'}$ and fourth term $f_{gbe}$ are forces due to the interactions between the dislocation and GBs, and the fifth term $f_H$ is the force due to the hydrogen atmosphere.

For the linear elastic analysis of plane problems, conventional approaches have been well established by Muskhelishvili [71], and later extended by Irwin [72] to obtain the well-known equations of the stress field near a sharp crack considering the first terms of series expansion. Creager and Paris [73] further derived the solutions for blunt cracks,

$$\begin{aligned}\sigma_{xx}^K = &\frac{K_I}{\sqrt{2\pi r}}\left(\cos\frac{\theta}{2} - \frac{1}{2}\sin\theta\sin\frac{3\theta}{2} - \frac{\rho}{2r}\cos\frac{3\theta}{2}\right)\\ &+ \frac{K_{II}}{\sqrt{2\pi r}}\left(-2\sin\frac{\theta}{2} - \frac{1}{2}\sin\theta\cos\frac{3\theta}{2} + \frac{\rho}{2r}\sin\frac{3\theta}{2}\right),\end{aligned} \tag{2}$$

$$\begin{aligned}\sigma_{yy}^K = &\frac{K_I}{\sqrt{2\pi r}}\left(\cos\frac{\theta}{2} + \frac{1}{2}\sin\theta\sin\frac{3\theta}{2} + \frac{\rho}{2r}\cos\frac{3\theta}{2}\right)\\ &+ \frac{K_{II}}{\sqrt{2\pi r}}\left(\frac{1}{2}\sin\theta\cos\frac{3\theta}{2} - \frac{\rho}{2r}\sin\frac{3\theta}{2}\right),\end{aligned} \tag{3}$$

$$\tau_{xy}^{K} = \frac{K_I}{\sqrt{2\pi r}}\left(\frac{1}{2}\sin\theta\cos\frac{3\theta}{2} - \frac{\rho}{2r}\sin\frac{3\theta}{2}\right) + \frac{K_{II}}{\sqrt{2\pi r}}\left(\cos\frac{\theta}{2} - \frac{1}{2}\sin\theta\sin\frac{3\theta}{2} - \frac{\rho}{2r}\cos\frac{3\theta}{2}\right), \tag{4}$$

where, $K_I$ and $K_{II}$ are the generalized mode-I and mode-II stress intensity factors (SIFs) assuming that the blunt notch is replaced by a sharp crack with a translation of the crack-tip on the curvature center $C$ [74], $(r, \theta)$ is defined as the polar coordinate $x_1O'y_1$ in **Fig. 1**. $f_K$ is then evaluated as $b\tau_{r\theta}^{K}$, where $b$ is the Burgers vector, and the shear component $\tau_{r\theta}^{K}$ in the polar coordinate system is calculated by the coordinate transformation as,

$$\begin{aligned}\tau_{r\theta}^{K} &= \cos\theta\sin\theta\left(\sigma_{yy}^{K} - \sigma_{xx}^{K}\right) + (\cos^2\theta - \sin^2\theta)\tau_{xy}^{K} \\ &= \frac{K_I}{2\sqrt{2\pi r}}\sin\frac{\theta}{2}\left(1 + \cos\theta + \frac{\rho}{r}\right) + \frac{K_{II}}{\sqrt{2\pi r}}\left(\cos\frac{3\theta}{2} + \frac{1}{2}\sin\theta\sin\frac{\theta}{2} - \frac{\rho}{2r}\cos\frac{\theta}{2}\right). \\ &= \frac{1}{2\sqrt{2\pi r}}\left[K_I\sin\frac{\theta}{2}\left(1 + \cos\theta + \frac{\rho}{r}\right) + K_{II}\cos\frac{\theta}{2}\left(-1 + 3\cos\theta - \frac{\rho}{r}\right)\right]\end{aligned} \tag{5}$$

The second term $f_{d^2}$ in the **Eq**.(1) is the self-image force caused by the free surface effect. Initially, to distinguish the stress components $\sigma_{yy}$ and $\tau_{yx}$ along the crack surface, image dislocation distributions are introduced along the crack surface [75]. The image dislocation ($\zeta_i$) distribution at a distance $\beta$ from the crack-tip, caused by the dislocation $\zeta$ emitted from either the crack-tip or the GB (see **Fig. 1**), is given by,

$$F_x(\beta) = \frac{-b}{\pi\rho_i}\sqrt{\frac{r}{|\beta|}}\left\{\cos\eta\cos\left(\phi - \frac{\theta}{2}\right) + \frac{1}{2}\sin\theta\sin\left(\phi - \eta + \frac{\theta}{2}\right) - \sin\phi\sin\left(2\phi - \eta - \frac{\theta}{2}\right)\right\}, \tag{6}$$

and

$$F_y(\beta) = \frac{-b}{\pi\rho_i}\sqrt{\frac{r}{|\beta|}}\left\{2\sin\eta\cos\left(\phi - \frac{\theta}{2}\right) + \cos\left(2\phi - \frac{\theta}{2}\right)\sin(\phi - \eta) - \frac{1}{2}\sin\theta\sin\left(\phi - \eta + \frac{\theta}{2}\right)\right\}. \tag{7}$$

Here $F_x(\beta)d\beta$ represents the sum of the $x$-directional components of the Burgers vector of the image dislocations between $\beta$ and $\beta + d\beta$, $b$ is the magnitude of the Burgers vector, and $\phi$ and $\eta$ are angles defined in **Fig. 1**. Substituting the **Eqs**.(6) and (7) into the stress field equations of a discrete dislocation [76] and integrating the functions over $\beta$ along the crack surface, the stress fields imposed by all image dislocations are given by,

$$\sigma_{xx}^{id} = \frac{\mu}{2\pi(1-\upsilon)}\left\{-\int_{-\infty}^{0}\frac{y[3(x-\beta)^2+y^2]}{[(x-\beta)^2+y^2]^2}F_x(\beta)\mathrm{d}\beta + \int_{-\infty}^{0}\frac{(x-\beta)[(x-\beta)^2-y^2]}{[(x-\beta)^2+y^2]^2}F_y(\beta)\mathrm{d}\beta\right\}, \tag{8}$$

$$\sigma_{yy}^{id} = \frac{\mu}{2\pi(1-\upsilon)}\left\{-\int_{-\infty}^{0} \frac{y[(x-\beta)^2-y^2]}{[(x-\beta)^2+y^2]^2} F_x(\beta)\mathrm{d}\beta + \int_{-\infty}^{0} \frac{(x-\beta)[(x-\beta)^2+3y^2]}{[(x-\beta)^2+y^2]^2} F_y(\beta)\mathrm{d}\beta\right\}, \tag{9}$$

and

$$\tau_{xy}^{id} = \frac{\mu}{2\pi(1-\upsilon)}\left\{-\int_{-\infty}^{0} \frac{(x-\beta)[(x-\beta)^2-y^2]}{[(x-\beta)^2+y^2]^2} F_x(\beta)\mathrm{d}\beta + \int_{-\infty}^{0} \frac{y[(x-\beta)^2-y^2]}{[(x-\beta)^2+y^2]^2} F_y(\beta)\mathrm{d}\beta\right\}. \tag{10}$$

where, the superscript '*id*' represents the image dislocation. Considering the practice of numerical solving, Shimokawa and Tsuboi [77] set the interval of the integration as -100 < $\beta$ < 0 nm, and calculated the $f_{d^2}$ in the same manner as $f_K$. Besides, some geometric constraints should be declared,

$$\eta = \arctan\left(\frac{r\sin\theta}{r\cos\theta-\alpha}\right), \tag{11}$$

$$\rho_r = \frac{r\sin\theta}{\sin\eta}, \tag{12}$$

$$\phi = \arctan\left(\frac{r\sin\theta}{r\cos\theta+\beta}\right), \tag{13}$$

and

$$\rho_i = \frac{r\sin\theta}{\sin\phi}, \tag{14}$$

where, the subscripts '$r$' and '$i$' in the **Eqs**.(12) and (14) represent real and imaginary (dislocations), respectively.

The third term $f_{dd'}$ in the **Eq**.(1) can be decomposed into the direct and indirect interaction force $f_{dd',d}$ and $f_{dd',i}$ caused by the residual dislocation $\zeta'$ after the dislocation $\zeta$ emission from the GB, as shown in **Fig. 1**. In other words, the residual dislocation $\zeta'$ exerts two types of stress fields to the dislocation $\zeta$: the direct and indirect stress fields. The direct force $f_{dd',d}$ can be evaluated as,

$$f_{dd',d} = -\frac{\mu b^2}{2\pi(1-\upsilon)\rho_r}. \tag{15}$$

On the other hand, the indirect force $f_{dd',i}$ can be calculated in the same manner as $f_{d^2}$, where the image dislocation distribution caused by the residual dislocation $\zeta'$ is obtained by setting $b = -b$, $\rho_i = \alpha + \beta$, $r = \alpha$, and $\phi = \eta = \theta = 0$. Then, substituting the distributions into the **Eqs**.(8)-(10), the value of $f_{dd',i}$ can be determined.

The fourth term $f_{gbe} = -4\Delta E_{gb}\rho_0/\pi(\rho_0^2 + \rho_r^2)$ (where, $\Delta E_{gb}$ is the energy change due to the dislocation emission, $\rho_0$ is the cutoff distance determined by the Rice-Thomson model [77]) is hypothesized to be caused by the local GB structural transition after dislocation emission from the GB, with details found in our previous study [78].

The fifth term $f_H$ is the shear force exerted on the dislocation induced by the hydrogen atmosphere [10],

$$f_H = \tau_H b \ , \tag{16}$$

Revisit **Fig. 1** and consider a Cartesian coordinate system $x_2O''y_2$ centered at the core of the emitted dislocation, and a single hydrogen dilatation line at a point with the polar coordinate ($r'$, $\varphi$). The shear stress exerted at the core of the emitted dislocation along the slip plane by the hydrogen dilatation line is [33],

$$\frac{\sigma_{r'r'}-\sigma_{\varphi\varphi}}{2}\sin2\varphi = -\frac{\mu\Delta a}{\pi r'^2}\sin2\varphi \tag{17}$$

where, $\sigma_{r'r'}$ and $\sigma_{\varphi\varphi}$ are nonzero stress components of the plane strain axisymmetric field at a distance $r'$ from the dilatation line, while $\Delta a = \Delta a'/2(1-v)$ with the "unconstrained area of expansion" $\Delta a' = V_H/N_A h$, where $V_H$ $(= N_A\Omega)$ is the partial molar volume of hydrogen in solution and $h$ is the distance between two successive hydrogen atoms along the dilatation line [33]. By applying the principle of linear superposition, the shear stress $\mathrm{d}\tau_H$ due to dilatation lines in an infinitesimal area d$S$ at the coordinate ($r'$, $\varphi$) is written as,

$$\mathrm{d}\tau_H = n\mathrm{d}S\left(-\frac{\mu\Delta a}{\pi r'^2}\right)\sin2\varphi \tag{18}$$

where, $n$ $(= Ch = N_L c_H h$, where, $C$ is the hydrogen concentration measured in atoms of hydrogen per unit lattice volume, and $N_L = 2/a_0^3$ denotes the number of solvent lattice atoms per unit lattice volume, where, $a_0$ is the lattice constant of the Fe matrix [33]) is the in-plane concentration of dilatation lines (also denoting the number of hydrogen atoms per unit area in the plane normal to the dilatation line), with $c_H$ is the hydrogen concentration near the dislocation core, treated as the primary thermodynamic state variable governing the local lattice resistance [15],

$$c_H(r',\varphi) = \frac{c_0}{1-c_0}\exp\left(-\frac{W_{int}}{k_B T}\right)\Big/\left(1 + \frac{c_0}{1-c_0}\exp\left(-\frac{W_{int}}{k_B T}\right)\right), \tag{19}$$

where, $W_{int}$ is the interaction energy between the hydrogen solute and the (total) hydrostatic stress field at this site,

$$W_{int} = -\Omega\left(\sigma_h^d + \sigma_h^K\right) , \tag{20}$$

where, $\sigma_h^d = \frac{\mu b(1+\upsilon)\sin\varphi}{3\pi(1-\upsilon)r'}$ is the hydrostatic component of the dislocation stress field itself, and $\sigma_h^K = \frac{\chi(\sigma_{xx}^K+\sigma_{yy}^K)}{3}$ is that of the crack-tip field ($\chi = 1$ for plane stress, and $1+\upsilon$ for plane strain, respectively), with the stress components $\sigma_{xx}^K$ and $\sigma_{yy}^K$ given by **Eqs**.(2)&(3). It should be noted here that the coordinate in $\sigma_h^K$ is not the $(r, \theta)$ pair for the position of the dislocation core, but the $(R, \Theta)$ pair for the position of a hydrogen solute, where, $R = \sqrt{r^2 + r'^2 + 2rr'\cos(\theta-\varphi)}$ and $\Theta = \text{atan2}(r\sin\theta + r'\sin\varphi, r\cos\theta + r'\cos\varphi)$. Integrating the above **Eq**.(18) over the entire area $S$ occupied by the hydrogen atmosphere, the hydrogen-induced net shear stress $\tau_H$ is,

$$\tau_H = -\frac{\mu\Omega N_L}{2\pi(1-\upsilon)}\int_{r_c}^{r_H}\int_0^{2\pi} c_H \frac{\sin 2\varphi}{r'} \mathrm{d}\varphi \mathrm{d}r' \tag{21}$$

where, $r_c$ is the core radius of the dislocation, $r_H$ is the radius of the hydrogen cluster around the dislocation. It should be noted here that, $c_H(r', \varphi)$ is the localized $H/M$ (hydrogen-to-metal) ratio around the emitted dislocation, while $c_0$ is the $H/M$ ratio in the crack-tip process zone rather than the equilibrium lattice concentration far from the crack-tip constrained by the Sieverts' law. As demonstrated by Song and Curtin [79], the hydrogen concentration in the near-tip region could reach 0.6~0.8 hydrogen per Fe atom.

While the stress field near the crack-tip under hydrogen environment has been configured above, the 2D Peierls model developed by Rice [7] and Rice and Beltz [8] (*i.e.*, the ***PRB*** framework) is introduced to describe the temperature-dependent activation energy barriers for dislocation nucleation,

$$U[\vec{\delta}(r)] = U_0 + \int_0^\infty \Phi[\vec{\delta}(r)]\mathrm{d}r + \frac{1}{2}\int_0^\infty s[\delta(r)]\cdot\delta(r)\mathrm{d}r - \int_0^\infty \frac{K_{II}^{eff}}{\sqrt{2\pi r}}\delta(r)\mathrm{d}r , \tag{22}$$

with

$$s[\delta(r)] = \frac{\mu}{2\pi(1-\upsilon)}\int_0^\infty \sqrt{\frac{\xi}{r}}\frac{\mathrm{d}\delta(\xi)/\mathrm{d}\xi}{r-\xi}\mathrm{d}\xi , \tag{23}$$

where, the first term $U_0$ is the elastic strain energy of the loaded cracked solid without any slip; the potential $\Phi$ is the change in atomic stacking energy due to a slip discontinuity $\vec{\delta}$; the third term accounts for the elastic interaction energy between the infinitesimal increments of slip; the fourth term represents the elastic interaction energy between the slip and the crack surface.

The activation energy per unit length under applied loading is the difference between the energies for the stable equilibrium slip distribution $\vec{\delta}_{stable}(r)$ and the saddle-point slip distribution $\vec{\delta}_{saddle}(r)$,

$$Q_{2d} = U[\vec{\delta}_{saddle}(r)] - U[\vec{\delta}_{stable}(r)] . \quad (24)$$

Proceeding via a perturbation analysis of the shear distribution, Rice and Beltz [8] obtained a closed-form approximation of the activation energy as,

$$\Theta_{2d} = \frac{(1-\upsilon)Q_{2d}}{\mu b^2} = m\left(1 - \sqrt{\frac{G}{\gamma_{usf}}}\right)^{3/2} , \quad (25)$$

where, the dimensionless factor $m$ is approximated as 0.287 due to the extremely weak dependence on $\gamma_{usf}/\mu b$ [80, 81], the applied energy release rate $G$ can be correlated with the effective mode-II SIF $K_{II}^{eff}$ via $G = (1-\upsilon)K_{II}^{eff^2}/2\mu$ [81].

By defining the localized mode-I SIF $K_I^*$ and mode-II SIF $K_{II}^*$ nominally (this is the key step from the continuum to atomistic mechanics, since atomistic details are considered), we can rewrite the total force in the hydrogen-informed Rice-Thomson-Shimokawa-Tsuboi (***RTST***) model as,

$$f_{tot} = b\tau_{r\theta}^* , \quad (26)$$

where, $\tau_{r\theta}^*$ is the nominal shear component within the polar coordinate system, and can be related to the nominal SIFs similar as **Eq**.(5),

$$\tau_{r\theta}^* = f_{tot}/b = \frac{1}{2\sqrt{2\pi r}}\left[K_I^*\sin\frac{\theta}{2}\left(1 + \cos\theta + \frac{\rho}{r}\right) + K_{II}^*\cos\frac{\theta}{2}\left(-1 + 3\cos\theta - \frac{\rho}{r}\right)\right] , \quad (27)$$

Meanwhile, the local strain energy density (*i.e.*, the first-order derivative of $U_0$ in **Eq**.(22) with respect to the volume) can be written as,

$$V_\varepsilon = \frac{1}{2}[\tilde{\sigma}]^T[\tilde{\varepsilon}] \quad (28)$$

with the stress and strain matrices are,

$$[\tilde{\sigma}] = \begin{matrix} \sigma_{xx}^* \\ \sigma_{yy}^* \\ \tau_{xy}^* \end{matrix}, [\tilde{\varepsilon}] = \begin{matrix} \varepsilon_{xx}^* \\ \varepsilon_{yy}^* \\ \gamma_{xy}^* \end{matrix} \quad (29)$$

with the strain matrix is,

$$[\tilde{\varepsilon}] = \begin{array}{l} \varepsilon_{xx}^* = \left(\sigma_{xx}^* - \upsilon\sigma_{yy}^*\right)/E \\ \varepsilon_{yy}^* = \left(\sigma_{yy}^* - \upsilon\sigma_{xx}^*\right)/E \\ \gamma_{xy}^* = 2(1+\upsilon)\tau_{xy}^*/E \end{array} \tag{30}$$

for plane stress, and,

$$[\tilde{\varepsilon}] = \begin{array}{l} \varepsilon_{xx}^* = (1+\upsilon)\left((1-\upsilon)\sigma_{xx}^* - \upsilon\sigma_{yy}^*\right)/E \\ \varepsilon_{yy}^* = (1+\upsilon)\left((1-\upsilon)\sigma_{yy}^* - \upsilon\sigma_{xx}^*\right)/E \\ \gamma_{xy}^* = 2(1+\upsilon)\tau_{xy}^*/E \end{array} \tag{31}$$

for plane strain. Explicitly, the components of the nominal stress field within the Cartesian coordinates are,

$$\begin{aligned} \sigma_{xx}^* = & \frac{K_I^*}{\sqrt{2\pi r}}\left(\cos\frac{\theta}{2} - \frac{1}{2}\sin\theta\sin\frac{3\theta}{2} - \frac{\rho}{2r}\cos\frac{3\theta}{2}\right) \\ & + \frac{K_{II}^*}{\sqrt{2\pi r}}\left(-2\sin\frac{\theta}{2} - \frac{1}{2}\sin\theta\cos\frac{3\theta}{2} + \frac{\rho}{2r}\sin\frac{3\theta}{2}\right) \end{aligned} \tag{32-a}$$

$$\begin{aligned} \sigma_{yy}^* = & \frac{K_I^*}{\sqrt{2\pi r}}\left(\cos\frac{\theta}{2} + \frac{1}{2}\sin\theta\sin\frac{3\theta}{2} + \frac{\rho}{2r}\cos\frac{3\theta}{2}\right) \\ & + \frac{K_{II}^*}{\sqrt{2\pi r}}\left(\frac{1}{2}\sin\theta\cos\frac{3\theta}{2} - \frac{\rho}{2r}\sin\frac{3\theta}{2}\right), \end{aligned} \tag{32-b}$$

$$\begin{aligned} \tau_{xy}^* = & \frac{K_I^*}{\sqrt{2\pi r}}\left(\frac{1}{2}\sin\theta\cos\frac{3\theta}{2} - \frac{\rho}{2r}\sin\frac{3\theta}{2}\right) \\ & + \frac{K_{II}^*}{\sqrt{2\pi r}}\left(\cos\frac{\theta}{2} - \frac{1}{2}\sin\theta\sin\frac{3\theta}{2} - \frac{\rho}{2r}\cos\frac{3\theta}{2}\right). \end{aligned} \tag{32-c}$$

By minimizing the strain energy density with respect to the nominal SIFs ($K_I^*$, $K_{II}^*$) under the linear constraint from shear stress (**Eq**.(27)), one can find values of two scalar variables, *i.e*. the nominal SIFs. Mathematically, this is a constrained optimization problem in finite-dimensional space, we can have the following solutions,

$$\begin{cases} K_I^* = \frac{D(A_{22}B - A_{12}C)}{A_{11}C^2 - 2A_{12}BC + A_{22}B^2} \\ K_{II}^* = \frac{D(A_{11}C - A_{12}B)}{A_{11}C^2 - 2A_{12}BC + A_{22}B^2} \end{cases} \tag{33}$$

where, the coefficients are,

$$\begin{aligned} B &= \sin(\theta/2)(1 + cos\theta + \rho/r) \\ C &= \cos(\theta/2)(-1 + 3cos\theta - \rho/r) \\ D &= 2\sqrt{2\pi r}(f_{tot}/b) \end{aligned} \tag{34}$$

The coefficients $A_{11}$, $A_{12}$, $A_{22}$ and more details about the derivation can be found in **Appendix A**. It should be noted that the nominal SIFs pair ($K_I^*$, $K_{II}^*$) is merely a scalar force-equivalent driving parameter rather than a reconstructed mixed-mode crack-tip field.

To involve the effect of GBs and hydrogen atmosphere on the dislocation nucleation, here we relate the effective mode-II SIF $K_{II}^{eff}$ in **Eq**.(22) with the above-defined localized mode-I SIF $K_I^*$ and localized mode-II SIF $K_{II}^*$ in **Eq**.(33) of the hydrogen-informed ***RTST*** model [7, 78],

$$K_{II}^{eff} = K_I^* \cos^2\frac{\theta}{2}\sin\frac{\theta}{2} + K_{II}^* \cos\frac{\theta}{2}\left(1 - 3\sin^2\frac{\theta}{2}\right), \tag{35-a}$$

Substituting **Eq**.(33) into **Eq**.(35-a), we have,

$$K_{II}^{eff} = D\beta_{eff} = \beta_{eff}\frac{2\sqrt{2\pi r}}{b} f_{tot} \tag{35-b}$$

where, the geometric coefficient is,

$$\beta_{eff} = \frac{(A_{22}B - A_{12}C)\cos^2\frac{\theta}{2}\sin\frac{\theta}{2} + (A_{11}C - A_{12}B)\cos\frac{\theta}{2}\left(1 - 3\sin^2\frac{\theta}{2}\right)}{A_{11}C^2 - 2A_{12}BC + A_{22}B^2} \tag{35-c}$$

Thus, the ***PRB*** framework now can be extended to consider the dislocation emission from a GB crack via,

$$\frac{(1-\upsilon)Q_{2d}}{\mu b^2} = m\left(1 - \sqrt{G'}\right)^{3/2}. \tag{36}$$

where, the normalized energy release rate is, $G' = \frac{(1-\upsilon)\left(K_{II}^{eff}\right)^2}{2\mu\gamma_{usf}} = \kappa f_{tot}^2$, with the coefficient is $\kappa = \frac{4\pi r(1-\upsilon)\beta_{eff}^2}{\mu\gamma_{usf} b^2}$. Inspired by previous studies [82, 83], and considering the temperature dependence, the activation Gibbs free energy can be rewritten as,

$$Q_{2d}(K_I, K_{II}, T) = (1 - T/T_m) Q_{2d}(K_I, K_{II}), \tag{37}$$

where, $T_m$ is the surface disordering temperature (to a first approximation, can be taken as the melting temperature) [82], and the activation enthalpy under zero temperature $Q_{2d}(K_I, K_{II}, T = 0)$ is given by,

$$Q_{2d}(K_I, K_{II}) = \frac{m\mu b^2}{1-\upsilon}\left(1 - \sqrt{\kappa} f_{tot}\right)^{3/2}, \tag{38}$$

Now, we introduce the reduced (or equivalent) SIF $K_r$ as a function of the applied Mode-I SIF $K_I$ and applied Mode-II SIF $K_{II}$ [84],

$$K_r \equiv K_r(K_I, K_{II}) = \sqrt{K_I^2 + K_{II}^2} \tag{39}$$

and define the (applied) mixity parameter $M_e = K_{II}/K_I$. It is noted that $K_r$ is indeed equivalent to the ***J***-integral ($\boldsymbol{J} = (K_I^2 + K_{II}^2)/E'$, where, $E' = E$ and $E/(1-\upsilon^2)$ for plane stress and plane strain, respectively). In practice, we use the normalized form of **Eq**.(38), *i.e.*, $\Theta_{2d}(K_I, K_{II}) = m\left(1 - \sqrt{\kappa} f_{tot}\right)^{3/2}$. The average rate of the dislocation nucleation can be described as,

$$\omega = \omega_0 N \exp\left(-\frac{Q_{3d}(K_I, K_{II}, T)}{k_B T}\right), \tag{40}$$

where, $\omega_0$ is the attempt frequency (to a first approximation, can be estimated as the Debye frequency, *i.e.*, $k_B T_D/\hbar$, where $T_D$ is the Debye temperature). Under the ***transition state theory*** (***TST***) framework, a dislocation nucleation event will occur once the reduced $K_r$ (of the applied mode-I SIF $K_I$ and mode-II SIF $K_{II}$) is equal to the most probable SIF $K_{Id}^p$,

$$\left.\frac{Q_{3d}(K_I, K_{II}, T)}{k_B T}\right|_{K_r = K_r^p} = \ln\left.\left(\frac{k_B T N \omega_0}{\dot{K}_r \Omega(K_I, K_{II}, T)}\right)\right|_{K_r = K_r^p}, \tag{41}$$

where, $N\omega_0$ represents the number of potential nucleation sites in the vicinity of the crack-tip multiplied by the attempt frequency $\omega_0$ [85]. The 3D energy barrier $Q_{3d}$ is estimated from the 2D energy barrier $Q_{2d}$ with a scaling factor $s_0$, *i.e.*, $Q_{3d} = s_0 Q_{2d}$. The activation volume-like term is defined as,

$$\Omega(K_I, K_{II}, T) = -\frac{\partial Q_{3d}}{\partial K_r} = \frac{3}{2}\left(1 - \frac{T}{T_m}\right)\frac{s_0 m \mu b^2}{1-\upsilon}\sqrt{\kappa\left(1 - \sqrt{\kappa} f_{tot}\right)}\frac{\mathrm{d} f_{tot}}{\mathrm{d} K_r}. \tag{42}$$

where, the derivative of the total force along the loading path, which is determined by holding the mixity parameter $M_e$ fixed, reads as,

$$\frac{d f_{tot}}{d K_r} = \left(\frac{\partial f_{tot}}{\partial K_I} + M_e \frac{\partial f_{tot}}{\partial K_{II}}\right) \Big/ \left(\frac{d K_r}{d K_I}\right) = \left.\frac{\partial f_{tot}}{\partial K_I}\right|_{loading} \Big/ \left(\frac{d K_r}{d K_I}\right) \tag{43}$$

where, $dK_r/dK_I = \sqrt{1 + M_e^2}$ is thus fixed by the definition of the applied mixed loading. Let us revisit the total force defined in **Eq**.(1), in which $f_K = f_K(K_I, K_{II})$ and $f_H = f_H(K_I, K_{II}; c_0)$ while other three terms in the r.h.s. are not functions of the applied load pairs $(K_I, K_{II})$, the remaining derivative in **Eq**.(43) is thus rewritten as,

$$\left.\frac{\partial f_{tot}}{\partial K_I}\right|_{loading} = \left.\frac{\partial f_K}{\partial K_I}\right|_{loading} + \left.\frac{\partial f_H}{\partial K_I}\right|_{loading} \tag{44}$$

Quoting the above defined coefficients, the contribution from the applied $K$-filed along the loading path is,

$$\left.\frac{\partial f_K}{\partial K_I}\right|_{loading}=\frac{\partial f_K}{\partial K_I}+M_e\frac{\partial f_K}{\partial K_{II}}=\frac{b(B+M_eC)}{2\sqrt{2\pi r}} \tag{45}$$

The hydrogen contribution is decomposed into two terms similarly,

$$\left.\frac{\partial f_H}{\partial K_I}\right|_{loading}=\frac{\partial f_H}{\partial K_I}+M_e\frac{\partial f_H}{\partial K_{II}} \tag{46}$$

with,

$$\frac{\partial f_H}{\partial K_j}=-\frac{b\mu\Omega N_L}{2\pi(1-\upsilon)}\iint c_H(1-c_H)u_j\frac{\sin 2\varphi}{r'}d\varphi dr',\ \ j=I,II \tag{47}$$

where, the partial derivatives $u_j$ are,

$$\begin{cases}u_I=\frac{2\Omega\chi}{3k_BT\sqrt{2\pi R}}\cos\frac{\Theta}{2}\\ u_{II}=\frac{-2\Omega\chi}{3k_BT\sqrt{2\pi R}}\sin\frac{\Theta}{2}\end{cases} \tag{48}$$

By numerically solving the **Eq**.(41) with the material constants listed in **Table 1**, the most probable mixed-mode SIF $K_r^p$ for dislocation nucleation from a GB crack can be obtained. Comparing the values of $K_r^p$ and the critical SIF required for intergranular crack cleavage $K_{IG}$ [78, 86], the fracture patterns can be determined as, *i.e.*, either the brittle cleavage or the ductile blunting via dislocation emission.

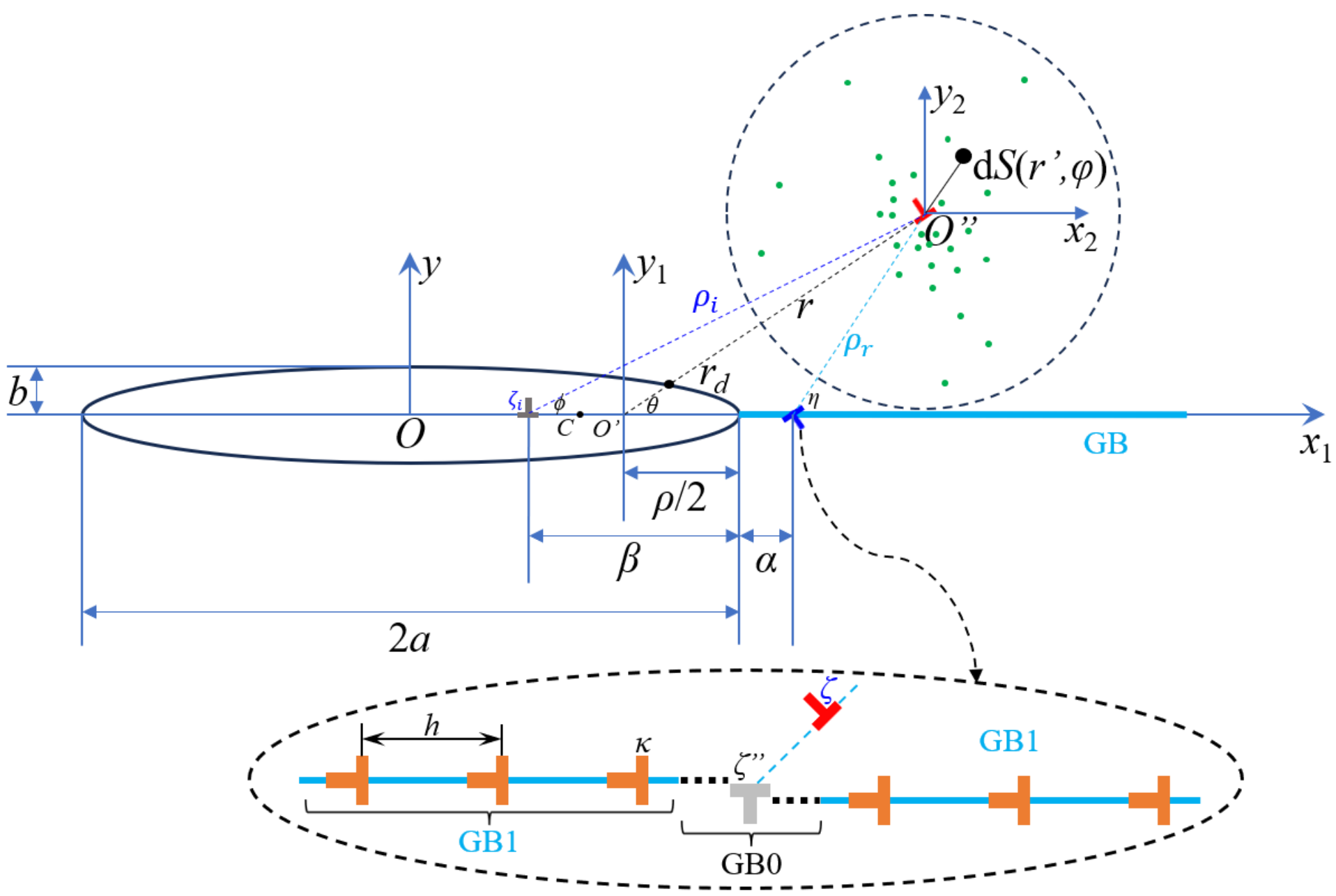

**Fig. 1**. Schematics of an edge dislocation emitted from the intergranular cack-tip under hydrogen environment. The green points represent hydrogen atoms.

### *2.2 Atomistic simulation protocols*

The essential knowledge required to advance the ***PRB*** model-based evaluation is the calculation of the generalized stacking fault energy (***GSFE***) curve, which should be a function of the hydrogen concentration [87]. Here I conduct the ***GSFE*** calculations of $\alpha$-Fe by using the Large-scale Atomic/Molecular Massively Parallel Simulator (LAMMPS) code [88]. The interatomic interaction between Fe and H atoms in α-Fe system is described by using the Finnis-Sinclair type modification [79] of the embedded-atom-method (EAM) potential originally developed by Ramasubramaniam *et al*. [89] and modified by Song and Curtin [79]. The time step for the velocity-Verlet integration is set as 1 fs. Periodic conditions are applied along *x*-, *y*- and *z*-axis directions. The simulation system of a cubic box is initially equilibrated at 300 K in the NPT ensemble. Subsequentially, hydrogen atoms are inserted in the simulation box randomly. The ***GSFE*** curve is evaluated by rigidly displacing two halves of the crystalline specimen, with the stacking sequence can be found in Ref.s [90, 91] for bcc lattice.

**Table 1**. Material parameters of bcc Fe [78].

| | | |
|---|---|---|
| Burgers vector, $b$ (Å) | 2.4825 | Ref.[8] |
| shear modulus, $\mu$ (GPa) | 69.3 | Ref.[8] |
| Poisson's ratio, $\upsilon$ | 0.291 | Ref.[8] |
| USF energy, $\gamma_{usf}$ (J/m$^2$) | 1.077 (Frenkel) | Ref.[7] |
| | 0.741 (EAM) | |
| Debye temperature, $T_D$ (K) | 477 (at 0 K) | Ref.[92] |
| | 373 (at 298 K) | Ref.[93] |
| surface disordering temperature, $T_m$ (K) | 1811 | Ref.[94] |

## 3. Results and discussions

### *3.1 Dislocation emission under hydrogen environment*

Let us define a mixity parameter $M_e = K_{II}/K_I$, **Fig. 2** shows the energetics of the dislocation emission from the crack tip under mixed-mode (I+II) loading. As shown in **Fig. 2**a, the critical $K_I$ at which the total force exerted on the dislocation is equal to 0, increases with the mixity. However, the energy maximums (*i.e.*, at the critical $K_I$) of each case are almost the same. This result demonstrates the self-consistency of the present theory, since the activation energy required for dislocation emission is not dependent on the loading patterns.

The material parameters of bcc Fe are listed in **Table 1**, the tip radius of a blunt crack is $\rho/b = 10$, and the distance $r = b$ is assumed for dislocation emission.

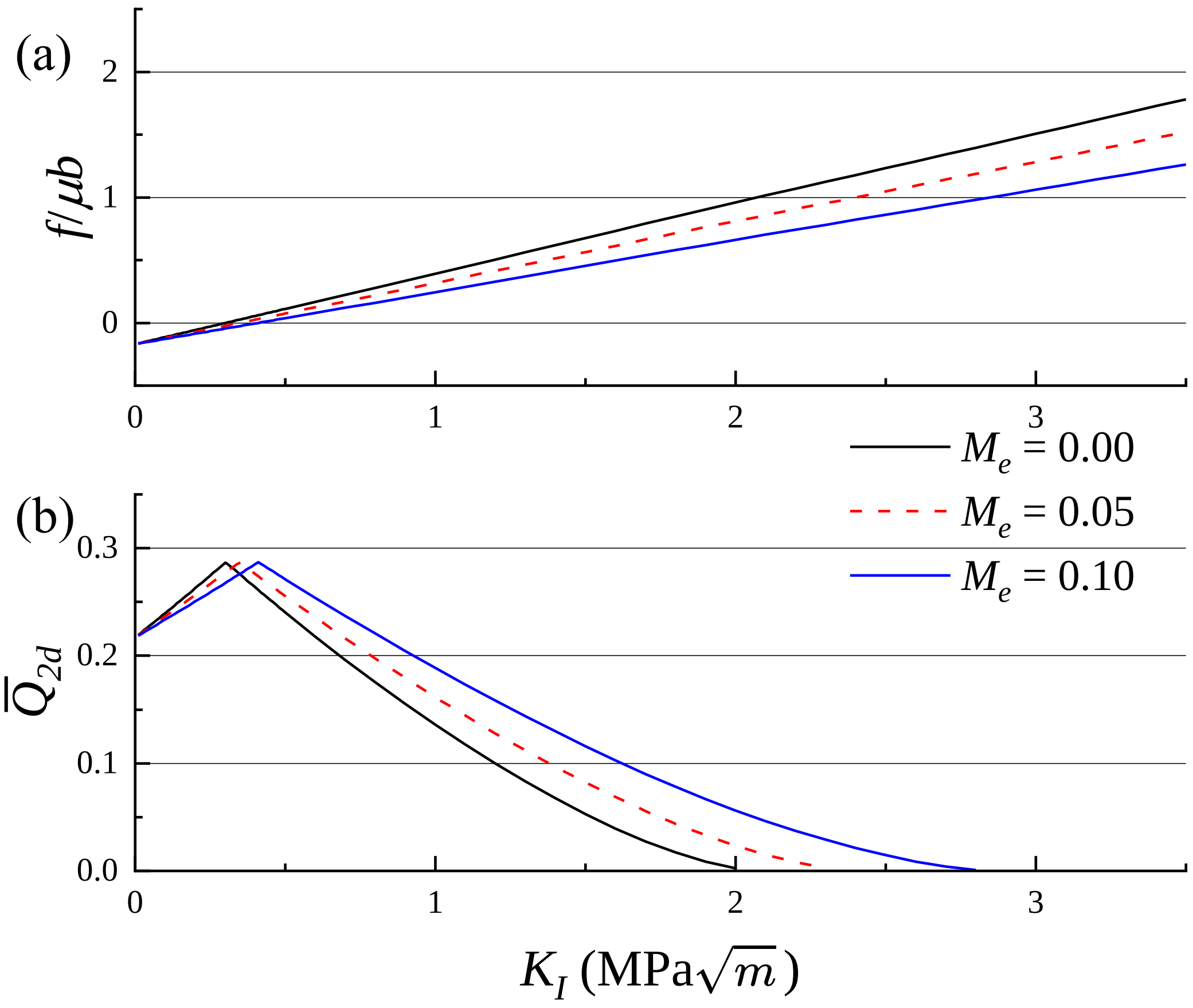

**Fig. 2**. Activation energy for mixed-mode (I+II) loading. (a) Normalized force $f/\mu b$ *v.s.* the applied mode-I SIF $K_I$; (b) Normalized activation energy $\tilde{Q}_{2d} = (1-\upsilon)Q_{2d}/\mu b^2$ *v.s.* $K_I$, where $\mu$ is the shear modulus, $b$ is the magnitude of the Burgers vector and $\upsilon$ is the Poisson's ratio. The hydrogen concentration ($H/M$ ratio) is set as $c_0$ = 100 appm, and the slip angle is $\theta$ = 30° here.

In order to access the rate dependence via the ***TST***, we further solve the most probable (reduced) SIF $K_r^p$. **Fig. 3** and **Fig. 4** show the slip-angle- and rate-dependence of $K_r^p$, respectively. While the $\theta$-dependence is evaluated for 3 different mixity values and 3 different loading rates in **Fig. 3**-(a), (b) & (c), these 9 cases are compared in **Fig. 3**-(d). The results reveal that the $\theta$ of $K_r^p$ is significantly influenced by the loading rate. It is found that there exist several local maximum and minimum values in the solution domain, and no solutions exist for $\theta$ = 180°. This is reasonable since $\theta$ = 180° corresponds to the position of fracture surfaces. It is also noted that even for $M_e = 0.00$, *i.e.*, the pure mode-I loading, the curve is not symmetric about the fracture plane (*i.e.*, $\theta$ = 180° or 0°). By revisiting the theoretical framework established in **§2.1**, this asymmetry is attributed to the presence of $f_{d^2}$ and $f_H$ terms.

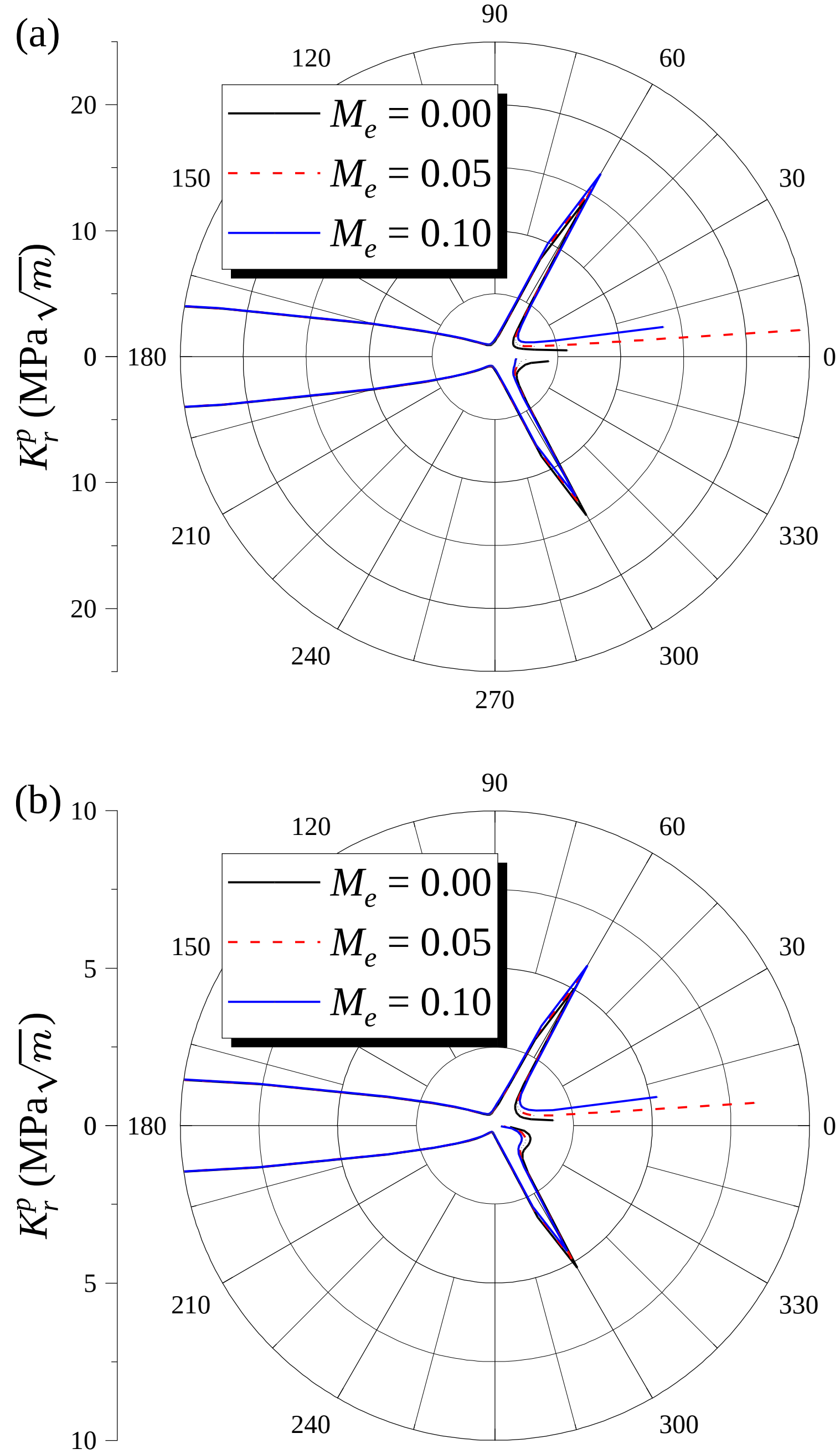

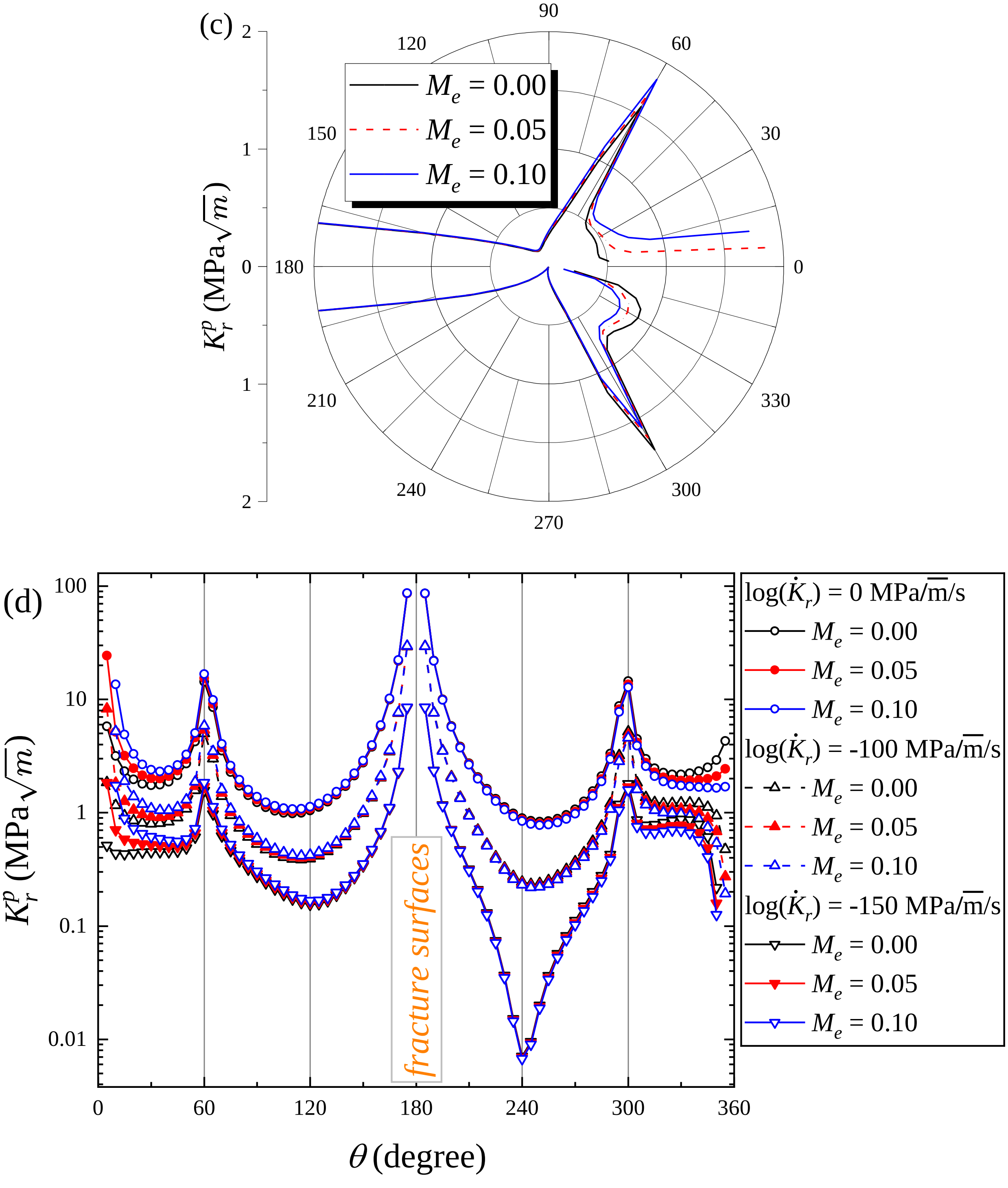

**Fig. 3**. The most probable (reduced) SIF $K_r^p$ as a function of the slip angle $\theta$ under various loading rates, (a) $\log(\dot{K}_r) = 0$ MPa$\sqrt{\text{m}}$/s; (b) $\log(\dot{K}_r) = -100$ MPa$\sqrt{\text{m}}$/s; (c) $\log(\dot{K}_r) = -150$ MPa$\sqrt{\text{m}}$/s. The hydrogen concentration is set as $c_0$ = 100 appm here. The panel-(d) compares all 9 cases presented in panels-(a), (b) & (c), with $\theta$ = 180° is marked as the position of fracture surfaces.

**Fig. 4** reveals that the rate-dependence of $K_r^p$ is significantly influenced by the mixity $M_e$. Specifically, the predicted SIF under high rates is ~ 1.75 MPa$\sqrt{\text{m}}$ for pure mode-I loading (*i.e.*, $M_e = 0.00$), thus in good agreement with the MD results in previous studies [79]. However, the reduced SIF $K_r^p$ seems to be not sensitive to the variation of $c_0$ within the studied range of [0, 2500] appm. As argued by Song and Curtin [79], a H-rich region with 0.6~0.8 H per Fe atom can be formed ahead of the crack-tip. Revisit the theoretical framework established in **§2.1**, the $H/M$ ratio is equivalent to the hydrogen concentration $c_0$ via $H/M = c_0 \times 10^{-6}$. We thus further study the relation of $K_r^p$ *v.s.* $H/M$ within a wider range of $H/M \in [0, 0.8]$ in **Fig. 5**.

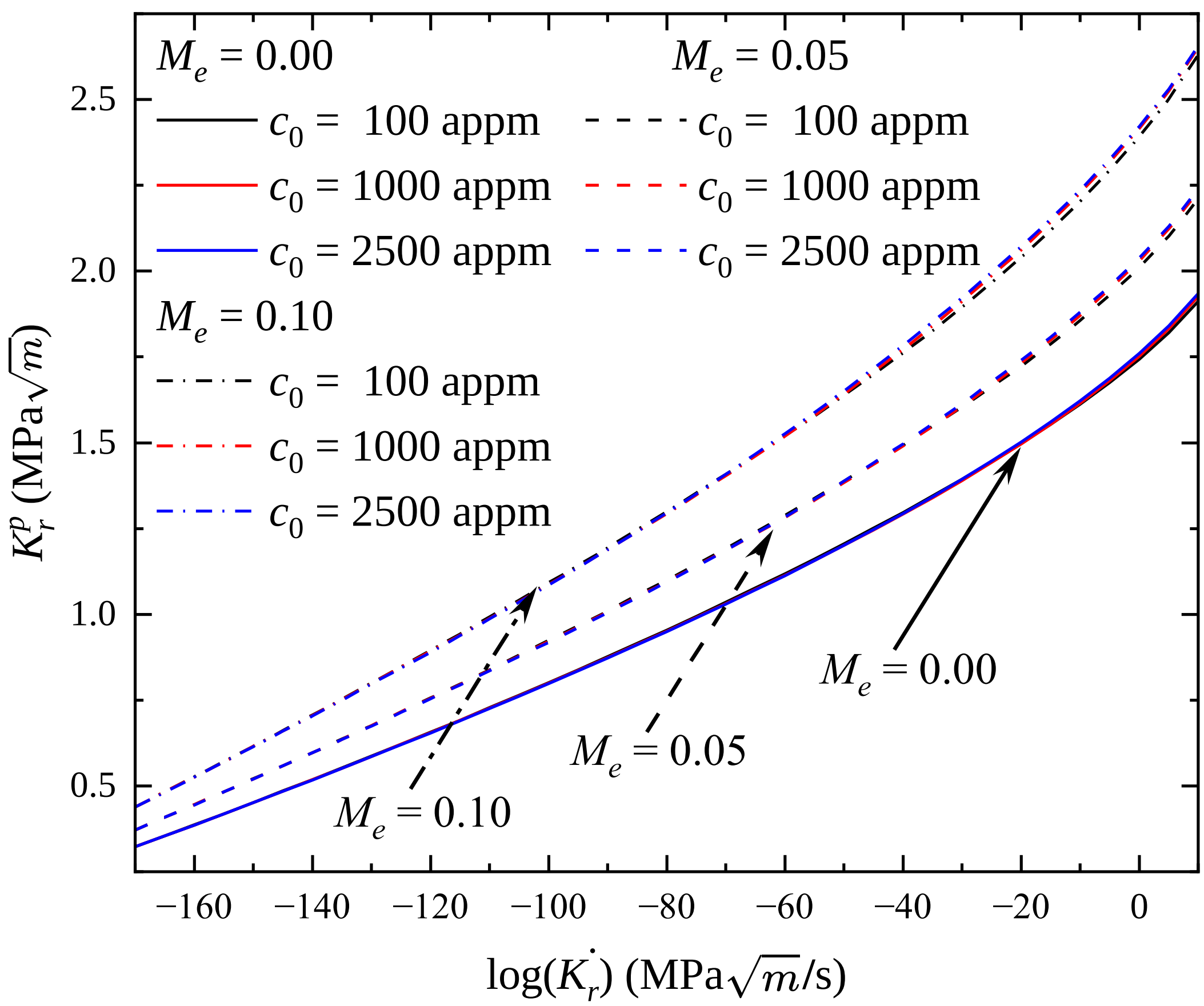

**Fig. 4**. The most probable (reduced) SIF $K_r^p$ as a function of the loading rate under different hydrogen concentrations ($c_0$ = 100, 1000 and 2500 appm). The slip angle is set as $\theta = 30°$ here.

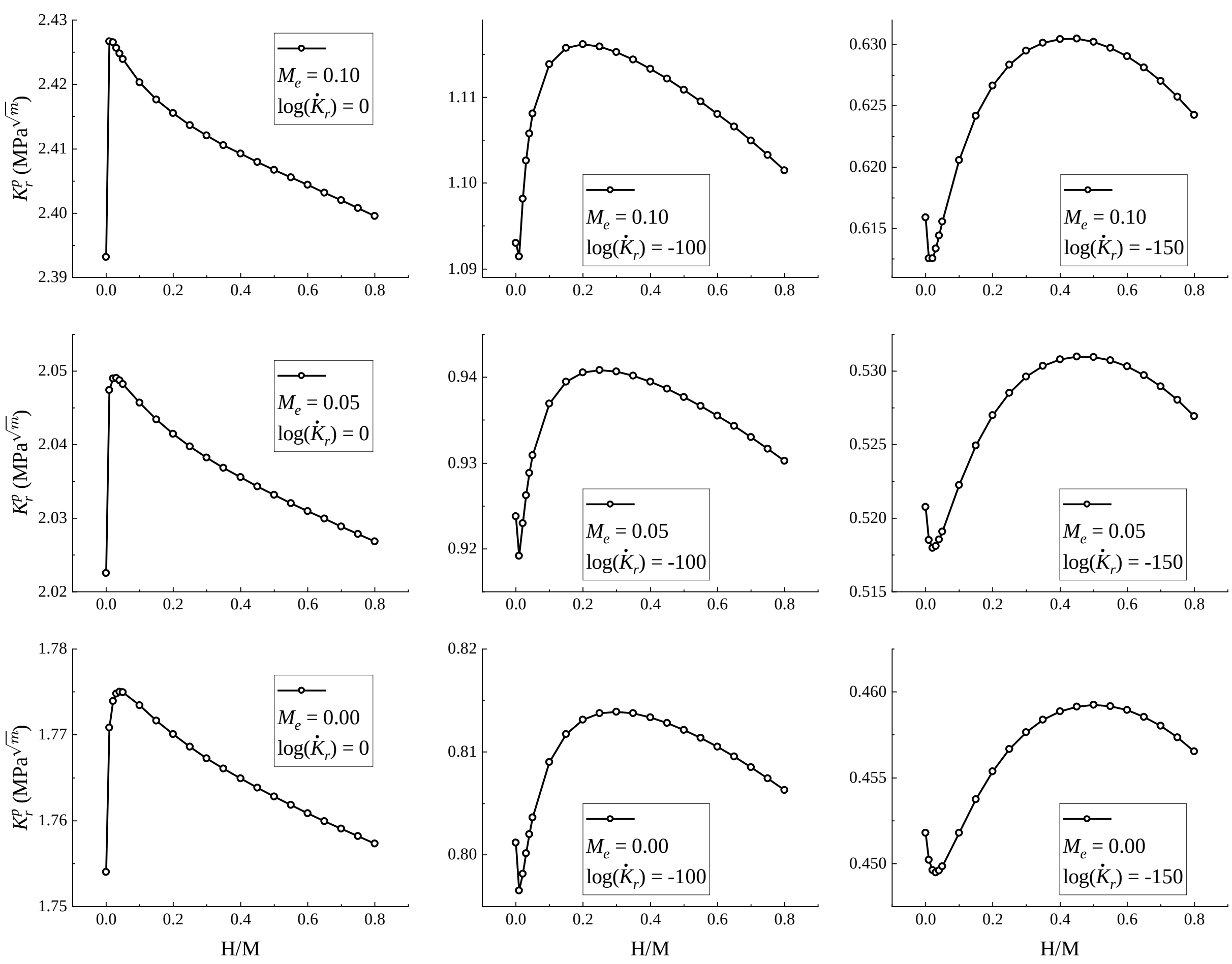

**Fig. 5**. The most probable (reduced) SIF $K_r^p$ as a function of the crack-tip hydrogen concentration $c_0$ (*i.e.*, the H/M ratio) for various loading rates ($\log(\dot{K}_r)$ = 0, −100, −150 MPa$\sqrt{\text{m}}$/s) under plane stress scenario. The slip angle is set as $\theta$ = 30° here.

As shown in **Fig. 5**, the non-monotonic relation of the most probable (reduced) SIF $K_r^p$ against the $H/M$ ratio (equivalent to $c_0$) is mainly determined by the loading rate while the mixity $M_e$ controls its amplitude. It is found that the hydrogen response of $K_r^p$ *v.s.* the hydrogen concentration $c_0$ can be roughly classified as two regimes for higher loading rate ($\log(\dot{K}_r) = 0$ MPa$\sqrt{\text{m}}$/s), *i.e.*, $K_r^p$ increases with $c_0$ in the lower concentration regime, but decreases with $c_0$ in the higher concentration regime. However, with the loading rate further decreases, $K_r^p$ would initially decrease with the $H/M$ ratio in a very narrow range, then changes with the $H/M$ ratio parabolically. While the dual role of hydrogen on the dislocation behavior has been reported recently [48], the present framework might shed a light to understand it theoretically. According to the definition in **§2.1**, and let $K_{II}^p = M_e K_I^p$, the most probable (applied) mode-I SIF $K_I^p$ is, $K_I^p = K_r^p/\sqrt{1+M_e^2}$. Thus, the most probable mode-I SIF $K_I^p$ required for dislocation emission is simply scaled with the reduced SIF $K_r^p$. Up to present, we are discussing the dislocation emission under plane stress sense. **Fig. A1** (see **Appendix A**) shows the results under plane strain scenario, which is almost the same as the plane stress counterpart only with slight difference of $K_r^p$ values.

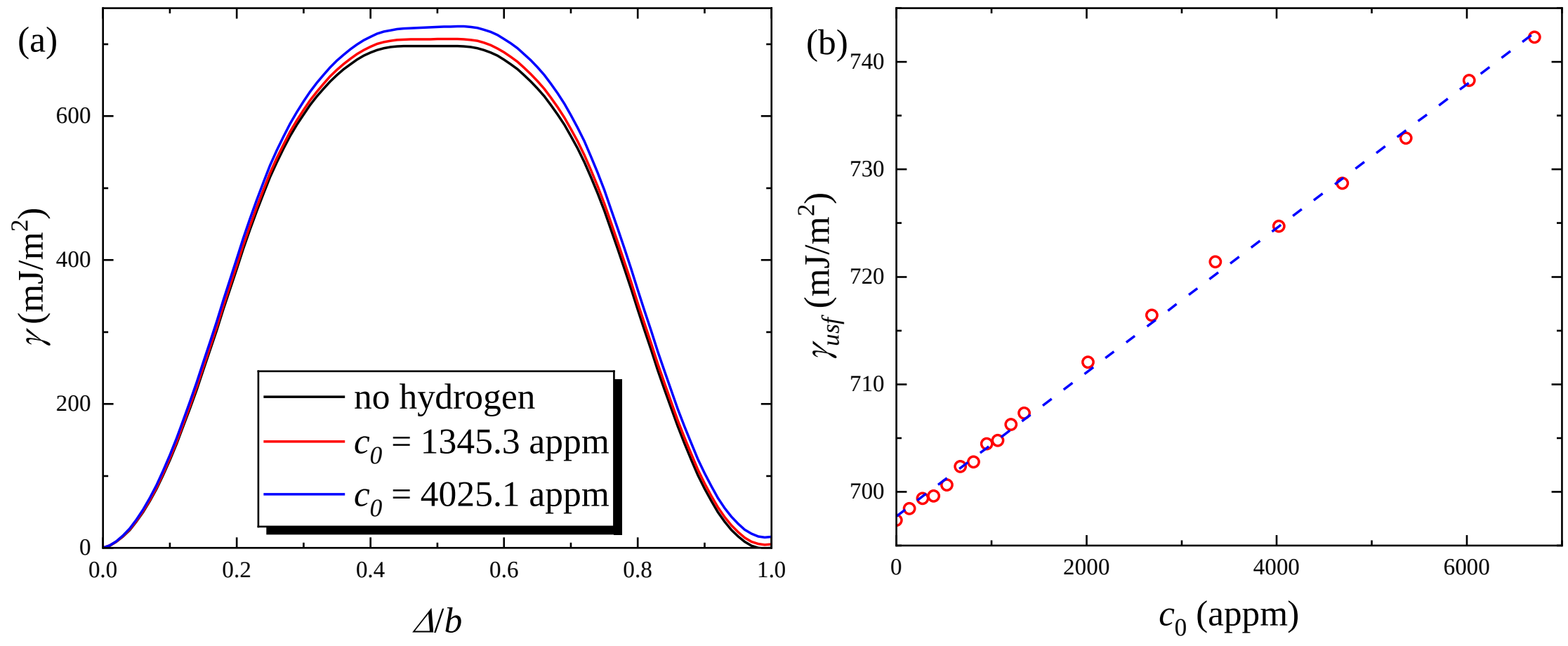

**Fig. 6**. (a) ***GSFE*** curves of the bcc Fe under various hydrogen concentrations, $c_0$ = 0, 1345.3, and 4025.1 appm, $b/a = \langle 111 \rangle/2$; (b) $\gamma_{usf}$ as a function of the hydrogen concentration $c_0$, the blue dash line is fitted to the red open circles linearly with the slope $k$ = 0.0067 (mJ/m$^2$)/appm.

Although sharp transitions are observed among different hydrogen concentration regimes in **Fig. 5**, the absolute value of $K_r^p$ does not vary significantly. Such small variations cannot explain the hydrogen-induced reduction of fracture toughness reported in literatures [10, 79]. It is noted that all above results are evaluated by hypothesizing the ***USF*** energy $\gamma_{usf}$ is not influenced by the hydrogen charging (see **Table 1**). However, the atomistic simulations demonstrate that $\gamma_{usf}$ might be significantly influenced by the hydrogen concentration $c_0$ in metals. **Fig. 6**-(a) shows the ***GSFE*** of the $\{112\}\langle 111 \rangle$ slip system of the bcc Fe, with the Burgers vector $b = a\langle 111 \rangle/2$, where $a$ is the lattice constant of Fe. **Fig. 6**-(b) shows that that the ***USF*** energy $\gamma_{usf}$ almost linearly increases with the increasing hydrogen concentration $c_0$. Assuming the linear correlation $\gamma_{usf} = \gamma_{usf}^0 + k \cdot c_0$ (where, $\gamma_{usf}^0$ is exactly the above-used ***USF*** energy of uncharged specimens in **Table 1**, and $k$ is the fitting parameter), and without loss of generality, we conduct parametric study of the slope $k$, as shown in **Fig. 7** for plane stress sense. It is found that for $M_e = 0.00$ scenario, the most probable SIF $K_r^p$ would increase with increasing ***USF*** energy slope $k$. The results also show that $K_r^p$ would increase with the hydrogen concentration, thus indicating the dislocation emission is suppressed by hydrogen charging, consistent with previous studies [10, 79].

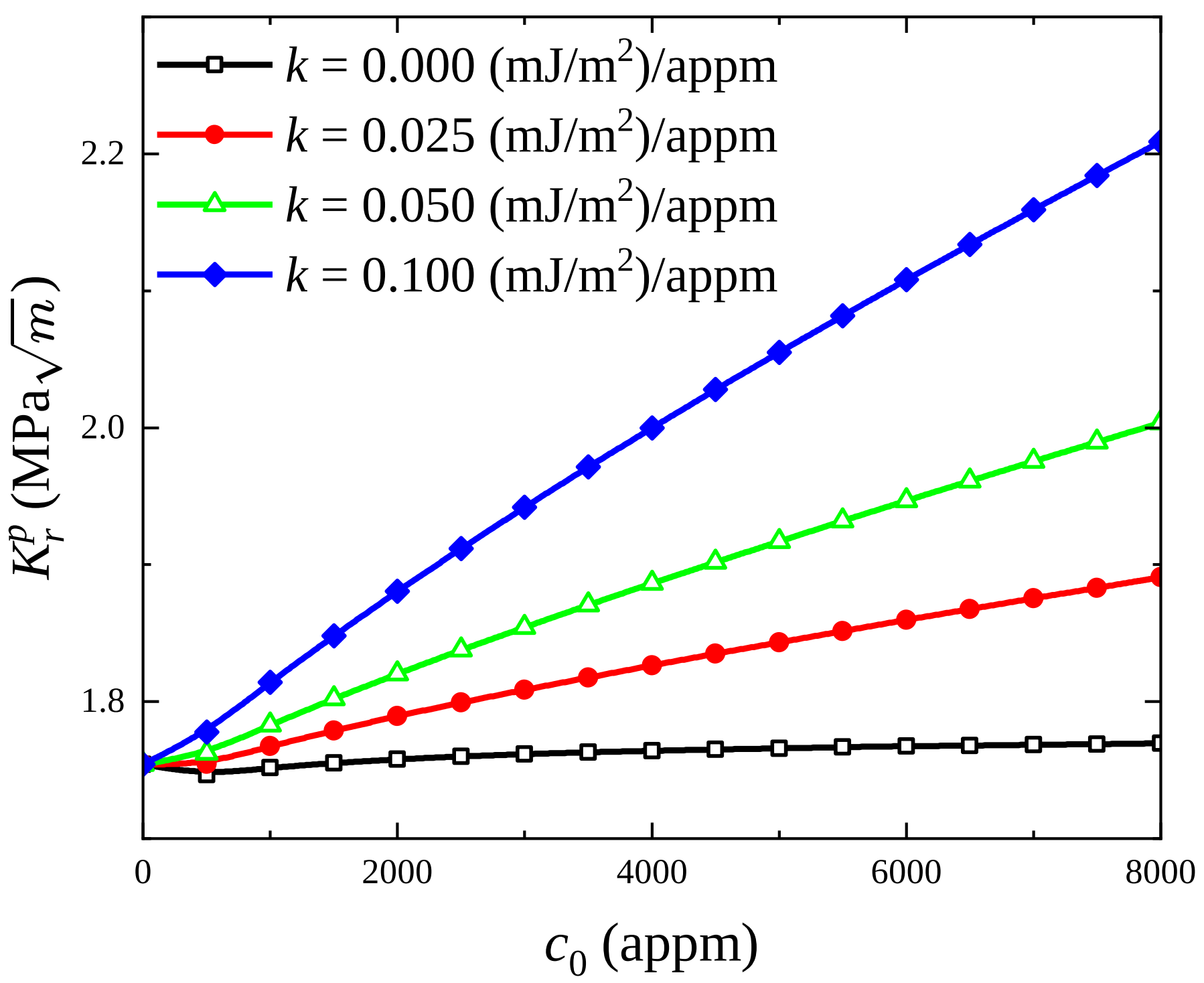

**Fig. 7**. The most probable (reduced) SIF $K_r^p$ as a function of the crack-tip hydrogen concentration $c_0$ for different ***USF*** energy slopes $k$ under plane stress scenario with the mixity $M_e = 0.00$, the loading rate $\log(\dot{K}_r) = 0$ MPa$\sqrt{\text{m}}$/s and the slip angle set as $\theta$ = 30°.

### *3.2 Effects of hydrogen atmosphere on the core structure of dislocation*

It is noted that the application of **Eq**.(19) and **Eq**.(21) actually indicates that the thermodynamic equilibrium is hypothesized in the present model, while the kinematic transportation of hydrogen atoms is not considered. However, for the dislocation nucleation within the characteristic time period of lattice vibration, it is adequate to assume the hydrogen is already distributed in the equilibrium status in the immediate vicinity of the crack tip. Nonetheless, it would be not enough to employ an equilibrium hydrogen distribution when the time scale is on the order of the characteristic time required for hydrogen diffusion, considering the strong singularity within the dislocation core region, and the effect of hydrogen atmosphere on the core structure [27, 28, 95]. Shishvan *et al.* [52] have demonstrated such separation of time scales for hydrogen enhanced cleavage, where the fast crack growth is supplied by the cavity gas rather than the bulk diffusion. Inspired by this time-scale separation argument, we consider the stochastic hydrogen occupancy when the loading outpaces the redistribution of hydrogen atmosphere. Let us revisit the seminal work by Rice and coworkers [7, 8, 96], and consider a simi-infinite crack lying on the plane $y = 0$ for $x < 0$ (see **Fig. 8**), and we now analyze the formation of slip (dislocation emission) on the plane $y = 0$ for $x > 0$, with defining the slip discontinuity distribution $\vec{\zeta}(x) = \vec{u}_x(x, 0^+) - \vec{u}_x(x, 0^-)$, lattice misfit potential energy $\Phi(\vec{\zeta})$, and lattice restoring stress $\tau[\zeta(x)] \equiv \Phi'[\zeta(x)] = \frac{\mathrm{d}\Phi}{\mathrm{d}\zeta}$. The stress $\tau_{xy}$ at any point $x$ on the slip plane arises from two sources, the external load $K$ and the internal dislocation distribution $b(x)$. Thus, we have, a) the external load stress: $\sigma_{load}(x) = \frac{K_{II}}{\sqrt{2\pi x}}$, is the stress at distance $x$ ahead of the tip for a mode-II crack loaded by the SIF $K_{II}$; b) the elastic back-stress: $\sigma_{self}(x;\zeta) = \frac{\mu}{2\pi(1-\upsilon)} \int_0^\infty \sqrt{\frac{x'}{x}} \frac{\mathrm{d}\zeta(x')/\mathrm{d}x'}{x-x'} \mathrm{d}x'$. In an infinite medium, the stress at $x$ caused by a dislocation at $x'$, is ~ $1/(x-x')$. However, near a crack, the term $\sqrt{x'/x}$ has to be introduced to account for the traction-free boundary condition on the crack surfaces ($x < 0$).

Considering the hydrogen (solute atoms) distribution as the decoration of the misfit potential similar as **Eq**.(22), the total energy functional equal to the sum of the misfit energy and the elastic potential energy minus the work of the load can be recast as,

$$E[\vec{\zeta}] = E_0 + \int_0^\infty \{\Phi[\vec{\zeta}(x)] + V_{sol}[\vec{\zeta}(x); c_H, \cdots]\}\mathrm{d}x + E_{elastic}[\vec{\zeta}] - W_{load}[\vec{\zeta}] \tag{49}$$

where, the first term $E_0$ (equivalent to the first term $U_0$ in **Eq**.(22)) is the energy of the loaded elastic solid in which $\zeta$ is constrained to be 0 along the slit and shear stress ahead of the crack is $\tau_0 = K_{II}/\sqrt{2\pi x}$.

The second term is the solute-decorated misfit energy, in which the original ***PN*** potential $\Phi$ is related but not equal, to the generalized stacking fault (***GSF***) interplanar potential $\Psi$, where $\Phi(\zeta) = \Psi(\Delta) - h\tau^2/2\mu$ with $\zeta$ is the magnitude of $\vec{\zeta}(x)$, $h$ is the interplanar spacing along the direction perpendicular to the slip plane, and $\tau$ ($= \mathrm{d}\Phi/\mathrm{d}\zeta = \mathrm{d}\Psi/\mathrm{d}\Delta$) is the shear stress (*i.e.*, the lattice restoring stress) on the slip plane in the direction of $\vec{\zeta}$. Inspired by recent studies [30, 97], where the so-called ***sPN*** model was proposed, the classical ***PN*** model is modified by scaling the interlayer potential $\Phi$ with a random variable $\omega$, to treat the stochastic occupancy of solute atoms within the core width,

$$\begin{aligned} E_{misfit} &= \int_0^\infty \{\Phi[\vec{\zeta}(x)] + V_{sol}[\vec{\zeta}(x); c_H, \cdots]\}\mathrm{d}x \\ &= \int_0^\infty \omega(x)\Phi[\vec{\zeta}(x)]\mathrm{d}x \\ &\approx \omega \int_0^\infty \Phi[\vec{\zeta}(x)]\mathrm{d}x \end{aligned} \tag{50}$$

With the random-amplitude interlayer potential, the equilibrium disregistry function is,

$$\zeta(x) = \frac{b}{\pi}\arctan\left(\frac{x}{\lambda}\right) + \frac{b}{2} \tag{51}$$

where, the equilibrium dislocation core width $\lambda$ is,

$$\lambda = \frac{h}{2(1-\upsilon)\omega} = \frac{\lambda_0}{\omega} \tag{52}$$

where, $\lambda_0 = h/2(1-\upsilon)$ is the core width in the classical ***PN*** model.

The third and fourth terms are also equivalent to their counterparts in **Eq.**(22), but only decay into the presently simple case with the slip is limited on the $y = 0$ plane, where the third term $E_{elastic}$ is the non-local elastic self-energy of the dislocation pile-up,

$$\begin{aligned} E_{elastic} &= \frac{1}{2}\int_0^\infty \zeta(x)\sigma_{self}(x;\zeta)\mathrm{d}x \\ &= \frac{\mu}{4\pi(1-\upsilon)}\int_0^\infty \zeta(x)\left[P.V.\int_0^\infty \sqrt{\frac{x\prime}{x}}\frac{1}{x-x\prime}\frac{\mathrm{d}\zeta(x\prime)}{\mathrm{d}x\prime}\mathrm{d}x'\right]\mathrm{d}x \end{aligned} \tag{53}$$

And the fourth term $W_{load}$ is the work done by the external $K$-field,

$$W_{load} = \int_0^\infty \frac{K_{II}}{\sqrt{2\pi x}}\zeta(x)\mathrm{d}x \tag{54}$$

It is noted that instead of using the original Frenkel sinusoidal function [7], Warner and Curtin [81] proposed a newly analytic form of the ***GSF*** interplanar potential $\Psi(\Delta)$, which could be evaluated from the atomistic simulations to involve the effects of field variables, *e.g*. the hydrogen concentration $c_H$, temperature $T$, *i.e*., $\Psi(\Delta) \to \Psi(\Delta; c_H, T, \cdots)$ with the four extrema ($\gamma_{ssf}, \gamma_{usf}, \gamma_{stf}, \gamma_{utf}$) of the generalized stacking fault energy (***GSFE***) curve as functions of these field variables. Upon now, we could arrive at a newly stochastic Peierls-Rice-Beltz (***sPRB***) model, to deal with the dislocation emission from a crack tip under environment of solute atoms, considering the short-range fluctuation of solute concentration within the dislocation core region. However, we would not extend further discussions of the ***sPRB*** model here, and its details will be presented in another up-coming paper.

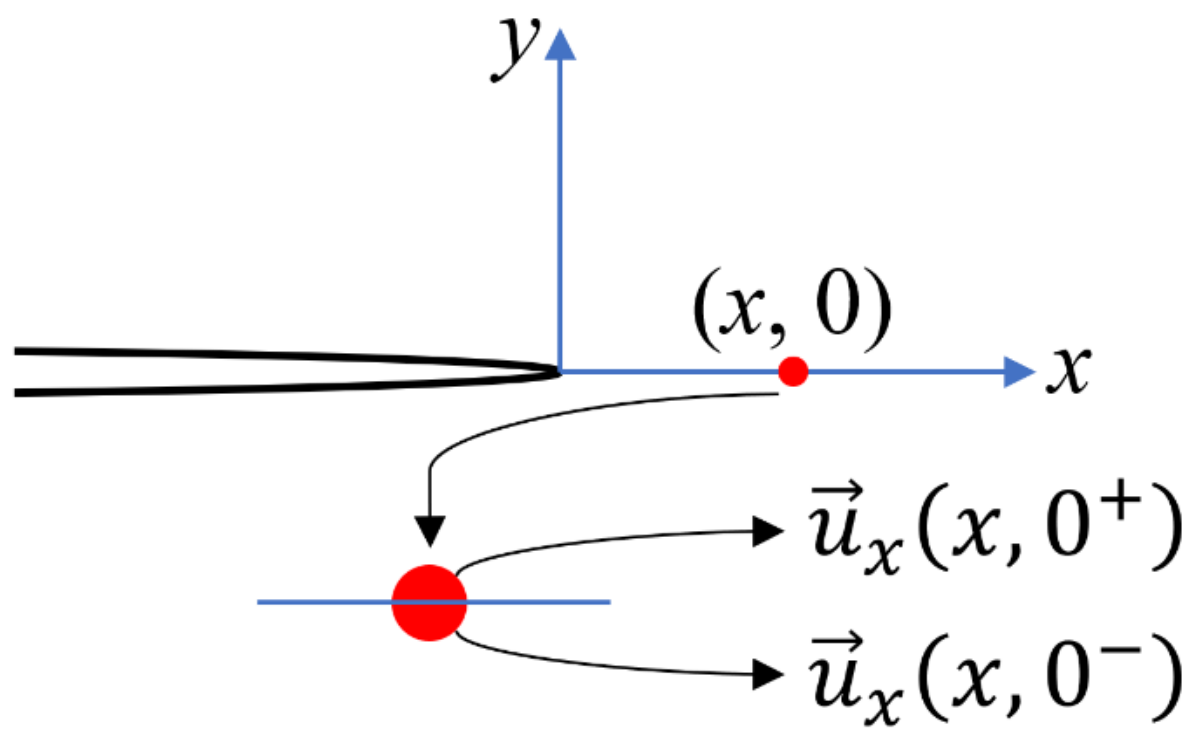

**Fig. 8**. Schematic configuration considered in Rice's theory. The red sphere represents a material point ($x$, 0) on the $x$-axis.

## 4. Conclusions

Stemming from the classical ***Rice-Beltz*** model, we have constructed an atomistically-informed theoretical framework to describe the crack-tip dislocation emission under hydrogen environments. The primary conclusions are as follows:

1. A rigorous theoretical model for predicting the critical SIF required for the initial dislocation emission under mixed-mode (I+II) loading is established, decoupling the nucleation event from macroscopic yielding and subsequent dislocation-forest interactions.

2. By introducing the minimization of the local strain energy density, we resolved the constraints of traditional pure mode-I (or mode-II) scenarios, determining the nominal SIFs ($K_I^*$ and $K_{II}^*$) for complex loading scenarios.

3. Hydrogen enters the model through the atmosphere shear force and the ***USF*** energy. The results show that the most probable SIF does not change with the hydrogen concentration monotonically, indicating a dual role of hydrogen on dislocation emission.

4. Recognizing the stochastic feature of the hydrogen solute distribution within the dislocation core width, this framework establishes the foundational boundary condition necessary for hierarchical ***HE*** modeling, paving the way for future integrations like the ***sPRB*** model.

### CRediT authorship contribution statement

**K. Zhao**: Conceptualization, Methodology, Software, Validation, Formal analysis, Investigation, Resources, Data Curation, Writing - Original Draft, Writing - Review & Editing, Visualization, Project administration, Funding acquisition.

### Declaration of competing interest

The authors declare that they have no known competing financial interests or personal relationships that could have appeared to influence the work reported in this paper.

### Acknowledgements

Financial supports provided by the National Natural Science Foundation of China (Grant No. 12102145), Natural Science Foundation of Jiangsu Province (Grant No. BK20210444) are acknowledged.

## Appendix A: Minimization of the strain energy density under mixed-mode loading

Revisit the **Eq**.(27) for the constraint from the shear stress equivalence, we can define $\eta = \rho/r$ for compactness, and let $s = \sin(\theta/2)$, $c = \cos(\theta/2)$. Using trigonometric identities, one can have,

$$\begin{aligned} \sin\theta &= 2sc \\ \cos\theta &= c^2 - s^2 \end{aligned} \tag{A.1}$$

The coefficients defined in **Eq**.(34) thus simplify to,

$$\begin{aligned} B &= s(1 + c^2 - s^2 + \eta) \\ C &= c(-1 + 3(c^2 - s^2) - \eta) \end{aligned} \tag{A.2}$$

The shear stress equivalence constraint **Eq**.(27) rearranges to,

$$BK_I^* + CK_{II}^* = D, \text{with}, D = 2\sqrt{2\pi r}f/b \tag{A.3}$$

The reduced stress fields are recast as,

$$\begin{cases} \sigma_{xx}^* = \frac{1}{\sqrt{2\pi r}}(K_I^* f_1 + K_{II}^* g_1) \\ \sigma_{yy}^* = \frac{1}{\sqrt{2\pi r}}(K_I^* f_2 + K_{II}^* g_2) \\ \tau_{xy}^* = \frac{1}{\sqrt{2\pi r}}(K_I^* f_3 + K_{II}^* g_3) \end{cases} \tag{A.4}$$

with using identities for multiple angles,

$$\begin{aligned} \sin(3\theta/2) &= 3s - 4s^3 \\ \cos(3\theta/2) &= 4c^3 - 3c \end{aligned} \tag{A.5}$$

The angular functions simplify to,

$$\begin{cases} f_1 = c - sc(3s - 4s^3) - \frac{\eta}{2}(4c^3 - 3c) = c\left(1 - 3s^2 + 4s^4 + \frac{3\eta}{2}\right) - 2\eta c^3 \\ f_2 = c + sc(3s - 4s^3) + \frac{\eta}{2}(4c^3 - 3c) = c\left(1 + 3s^2 - 4s^4 - \frac{3\eta}{2}\right) + 2\eta c^3 \\ f_3 = sc(4c^3 - 3c) - \frac{\eta}{2}(3s - 4s^3) = sc^2(4c^2 - 3) - \frac{\eta s}{2}(3 - 4s^2) \\ g_1 = -2s - sc(4c^3 - 3c) + \frac{\eta}{2}(3s - 4s^3) = -2s - sc^2(4c^2 - 3) + \frac{\eta s}{2}(3 - 4s^2) \\ g_2 = f_3 \\ g_3 = f_1 \end{cases} \tag{A.6}$$

These simplifications reduce higher powers and facilitate squaring in the energy expression.

The strain energy density is,

$$V_\varepsilon = \frac{1}{2E}\left[(\sigma_{xx}^*)^2 + \left(\sigma_{yy}^*\right)^2 - 2\upsilon\sigma_{xx}^*\sigma_{yy}^* + 2(1+\upsilon)\left(\tau_{xy}^*\right)^2\right] \quad \text{(A.7)}$$

for plane stress, and,

$$V_\varepsilon = \frac{1+\upsilon}{2E}\left[(1-\upsilon)(\sigma_{xx}^*)^2 + (1-\upsilon)\left(\sigma_{yy}^*\right)^2 - 2\upsilon\sigma_{xx}^*\sigma_{yy}^* + 2\left(\tau_{xy}^*\right)^2\right] \quad \text{(A.8)}$$

for plane strain.

Let $k = 1/\sqrt{2\pi r}$, substituting the stresses yields a quadratic form:

$$V_\varepsilon = \frac{k^2}{2E}[A_{11}(K_I^*)^2 + 2A_{12}K_I^*K_{II}^* + A_{22}(K_{II}^*)^2] \quad \text{(A.9)}$$

for plane stress, and,

$$V_\varepsilon = \frac{(1+\upsilon)k^2}{2E}[A'_{11}(K_I^*)^2 + 2A'_{12}K_I^*K_{II}^* + A'_{22}(K_{II}^*)^2] \quad \text{(A.10)}$$

for plane strain, where the coefficients, incorporating the simplified angular functions, are,

$$\begin{aligned} A_{11} &= f_1^2 + f_2^2 - 2\upsilon f_1 f_2 + 2(1+\upsilon)f_3^2 \\ A_{12} &= f_1 g_1 + f_2 g_2 - \upsilon(f_1 g_2 + f_2 g_1) + 2(1+\upsilon)f_3 g_3 \\ A_{22} &= g_1^2 + g_2^2 - 2\upsilon g_1 g_2 + 2(1+\upsilon)g_3^2 \end{aligned} \quad \text{(A.11)}$$

for plane stress, and,

$$\begin{aligned} A'_{11} &= (1-\upsilon)(f_1^2 + f_2^2) - 2\upsilon f_1 f_2 + 2f_3^2 \\ A'_{12} &= (1-\upsilon)(f_1 g_1 + f_2 g_2) - \upsilon(f_1 g_2 + f_2 g_1) + 2f_3 g_3 \\ A'_{22} &= (1-\upsilon)(g_1^2 + g_2^2) - 2\upsilon g_1 g_2 + 2g_3^2 \end{aligned} \quad \text{(A.12)}$$

for plane strain.

To minimize the quadratic $V_\varepsilon$ subject to the linear constraint, one can use the method of Lagrange multipliers or direct substitution. By defining the vector $\mathbf{K} = [K_I^*, K_{II}^*]^T$, quadratic matrix $\boldsymbol{Q} = \begin{bmatrix} A_{11} & A_{12} \\ A_{21} & A_{22} \end{bmatrix}$ ($A_{12} = A_{21}$), constraint vector $\boldsymbol{a} = [B, C]^T$, and scalar $D$, the objective is: $\min \frac{1}{2}\mathbf{K}^T\boldsymbol{Q}\mathbf{K}$ s.t. $\boldsymbol{a}^T\mathbf{K} = D$, and the closed-form solution is,

$$\mathbf{K} = \boldsymbol{Q}^{-1}\boldsymbol{a}\left(\frac{D}{\boldsymbol{a}^T\boldsymbol{Q}^{-1}\boldsymbol{a}}\right) \quad \text{(A.13)}$$

where, the inverse matrix is,

$$\boldsymbol{Q}^{-1} = \frac{1}{\det \boldsymbol{Q}}\begin{bmatrix} A_{22} & -A_{12} \\ -A_{21} & A_{11} \end{bmatrix} \quad \text{(A.14)}$$

with the determinant is,

$$\det \boldsymbol{Q} = A_{11}A_{22} - A_{12}^2 \tag{A.15}$$

Then,

$$\boldsymbol{a}^T\boldsymbol{Q}^{-1}\boldsymbol{a} = \frac{A_{11}C^2 - 2A_{12}BC + A_{22}B^2}{\det Q} \tag{A.16}$$

It is also noted that $V_\varepsilon$ is a positive definite quadratic form in the stress intensity factors, meaning that the Hessian (second order derivative matrix) is positive definite, guaranteeing a global minimum rather than a maximum or saddle point.

Alternatively, via substitution of $K_{II}^* = (D - BK_I^*)/C$ (or $K_I^* = (D - CK_{II}^*)/B$) into $V_\varepsilon$, and let $\mathrm{d}V_\varepsilon/\mathrm{d}K_I^* = 0$ (or $\mathrm{d}V_\varepsilon/\mathrm{d}K_{II}^* = 0$), one can have,

$$\begin{aligned} &A_{11}C^2K_I^* + A_{12}C(D - 2BK_I^*) + A_{22}B(-D + BK_I^*) = 0 \\ &\Rightarrow K_I^* = \frac{D(A_{22}B - A_{12}C)}{A_{11}C^2 - 2A_{12}BC + A_{22}B^2} \end{aligned} \tag{A.17}$$

or,

$$\begin{aligned} &A_{11}C(-D + CK_{II}^*) + A_{12}B(D - 2CK_{II}^*) + A_{22}B^2K_{II}^* = 0 \\ &\Rightarrow K_{II}^* = \frac{D(A_{11}C - A_{12}B)}{A_{11}C^2 - 2A_{12}BC + A_{22}B^2} \end{aligned} \tag{A.18}$$

Actually, the denominator in **Eq**.(A.17) & **Eq**.(A.18) $\Delta = A_{11}C^2 - 2A_{12}BC + A_{22}B^2$ represents $\boldsymbol{a}^T\boldsymbol{Q}^{-1}\boldsymbol{a} \cdot \det \boldsymbol{Q}$, ensuring the convexity (assuming $\boldsymbol{Q} > 0$).

Assuming the ***USF*** energy is independent of the local hydrogen concentration $c_H$, **Fig. A1** shows the most probable (reduced) SIF $K_r^p$ as a function of the environmental hydrogen concentration $c_0$ under plane strain, is almost the same as its plane stress counterpart shown in **Fig. 5**.

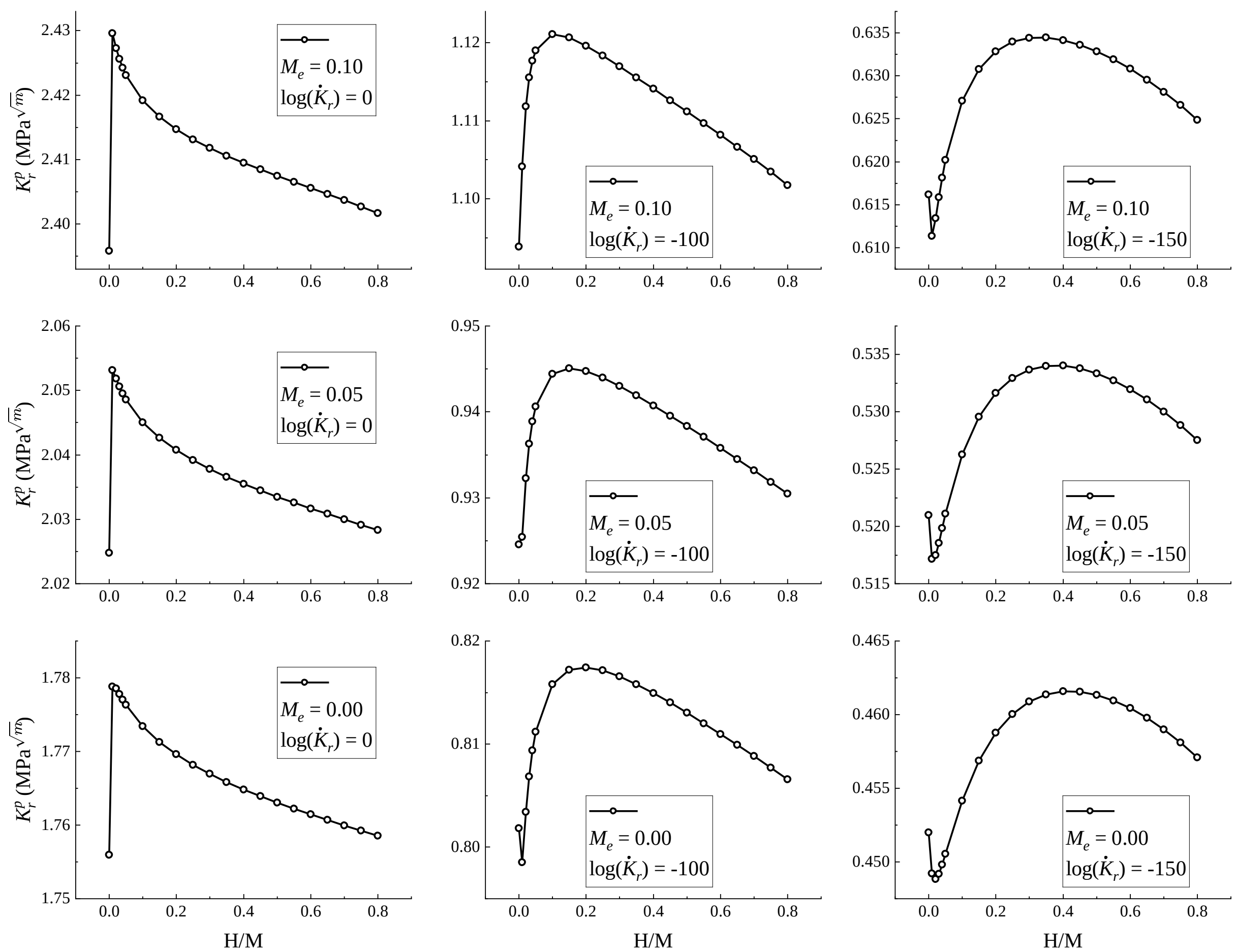

**Fig. A1**. The most probable (reduced) SIF $K_r^p$ as a function of the crack-tip hydrogen concentration $c_0$ (*i.e.*, the H/M ratio) for various loading rates ($\log(\dot{K}_r)$ = 0, -100, -150 MPa$\sqrt{\mathrm{m}}$/s) under plane strain scenario. The slip angle is set as $\theta$ = 30° here.

## Data availability statement

Data will be made available on reasonable request.